\documentclass[
    ,final            
  ]
  {aipproc}

\layoutstyle{6x9}

\begin{document}

\title{The bimodality of type Ia Supernovae}


\author{F. Mannucci}{
  address={INAF - IRA, Firenze, Italia}
}

\author{N. Panagia}{
  address={STScI, USA; INAF - OAC - Catania, Italia; SN Ltd - Virgin
  Gorda, BVI}
}

\author{M. Della Valle}{
  address={INAF - OAA - Firenze, Italia}
}

\begin{abstract}
We comment on the presence of a bimodality in the distribution
of delay time between the formation of the progenitors and their explosion as
type Ia SNe. Two "flavors" of such bimodality are present in the literature:
a {\em weak} bimodality, in which 
type Ia SNe must explode from both young and old progenitors, and a 
{\em strong}
bimodality, in which about half of the systems explode within 10$^8$ years
from formation. The {\em weak} bimodality is observationally based on
the dependence of the rates with the host galaxy Star Formation Rate (SFR), 
while the
{\em strong} one on the different rates in radio-loud and 
radio-quiet early-type galaxies. 
We review the evidence for these bimodalities. 
Finally, we estimate the fraction of SNe which are missed by optical and
near-IR searches because of dust extinction in massive starbursts.
\end{abstract}

\maketitle


\section{Introduction}

The supernova (SN) rates in different types of galaxies give strong
informations about the progenitors. For example, soon after the introduction
of the distinction between ``type I'' and ``type II'' SNe \citep{minkowski41},
\citet{vandenbergh59} pointed out that type IIs are frequent 
in late type galaxies ``which suggest their affiliation with Baade's
population I''. On the contrary, type Is, are 
the only type observed in elliptical galaxies and this fact "suggests that
they occur among old stars". This conclusion is still often accepted,
even if it is now known not to be generally valid: first, 
SN Ib/c were included in the broad class of ``type I'' SNe, and, second, 
also a
significant fraction of SNe Ia are known to have young progenitors.

\begin{figure}   
  \includegraphics[height=.31\textheight]{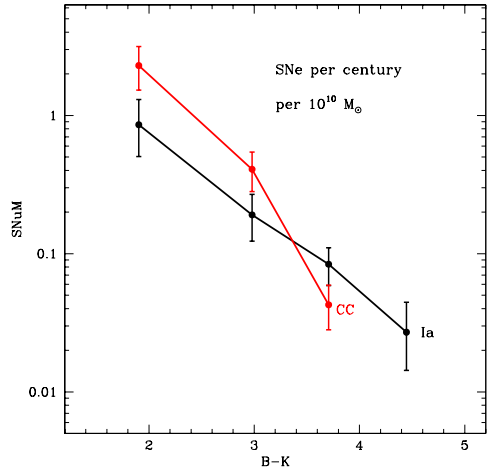}
  \caption{SN rate per unit stellar mass as a function of the B--K color of the
  parent galaxy (from \citet{mannucci05}) showing the strong increase of all
  the rates toward blue galaxies}
\end{figure}

\section{The {\em weak} bimodality in type Ia SNe}

In 1983, \citet{greggio83} showed that the canonical binary star 
models for type Ia SNe naturally predict that these systems
explode from progenitors of very different ages,
from a few 10$^7$ to 10$^{10}$ years. The strongest observational
evidence that this is the case
was provided by \citet{mannucci05} who analyzed the SN rate per unit stellar
mass in galaxies of all types. They found that the
bluest galaxies, hosting the highest Star Formation Rates (SFRs), 
have SN Ia rates about 30 times larger than those in the reddest, 
quiescent galaxies. 
The higher rates in actively star-forming galaxies imply that
a significant fraction of SNe must be due to young stars,
while SNe from old stellar populations are also
needed to reproduce the SN rate in quiescent galaxies. This lead
\citet{mannucci05} to introduce the simplified 
two component model for the SN Ia rate 
(a part proportional to the stellar mass and another part to the SFR).
These results were later confirmed by \citet{sullivan06}, while
\citet{scanna05}, \citet{matteucci06} and \citet{calura07}
successfully applied this model
to explain the chemical evolution of galaxies and galaxy clusters.
A more accurate description is based on the Delay Time Distribution (DTD),
which is found to span a wide range of delay time between a few $10^7$ 
to a few $10^{10}$ years (\citet{mannucci06}).
The presence of a strong observational result and the agreement 
with the predictions of several models (see also \citet{greggio05}) 
make this conclusion very robust.

\section{The {\em strong} bimodality in type Ia SNe}

\citet{dellavalle05} studied the dependence of the SN Ia rate 
in early-type galaxies on the radio power of the host galaxies, and
concluded that the higher rate observed in radio-loud galaxies
is due to minor episodes of accretion of gas or capture of small
galaxies. Such events result in both fueling the 
central black hole, producing the radio activity,
and in creating a new generation of stars, producing the increase in the
SN rate.
This effect can be used to derive information on the DTD of type Ia SNe
once a model of galaxy stellar population is introduced.

\begin{figure}   
  \includegraphics[height=6.3cm]{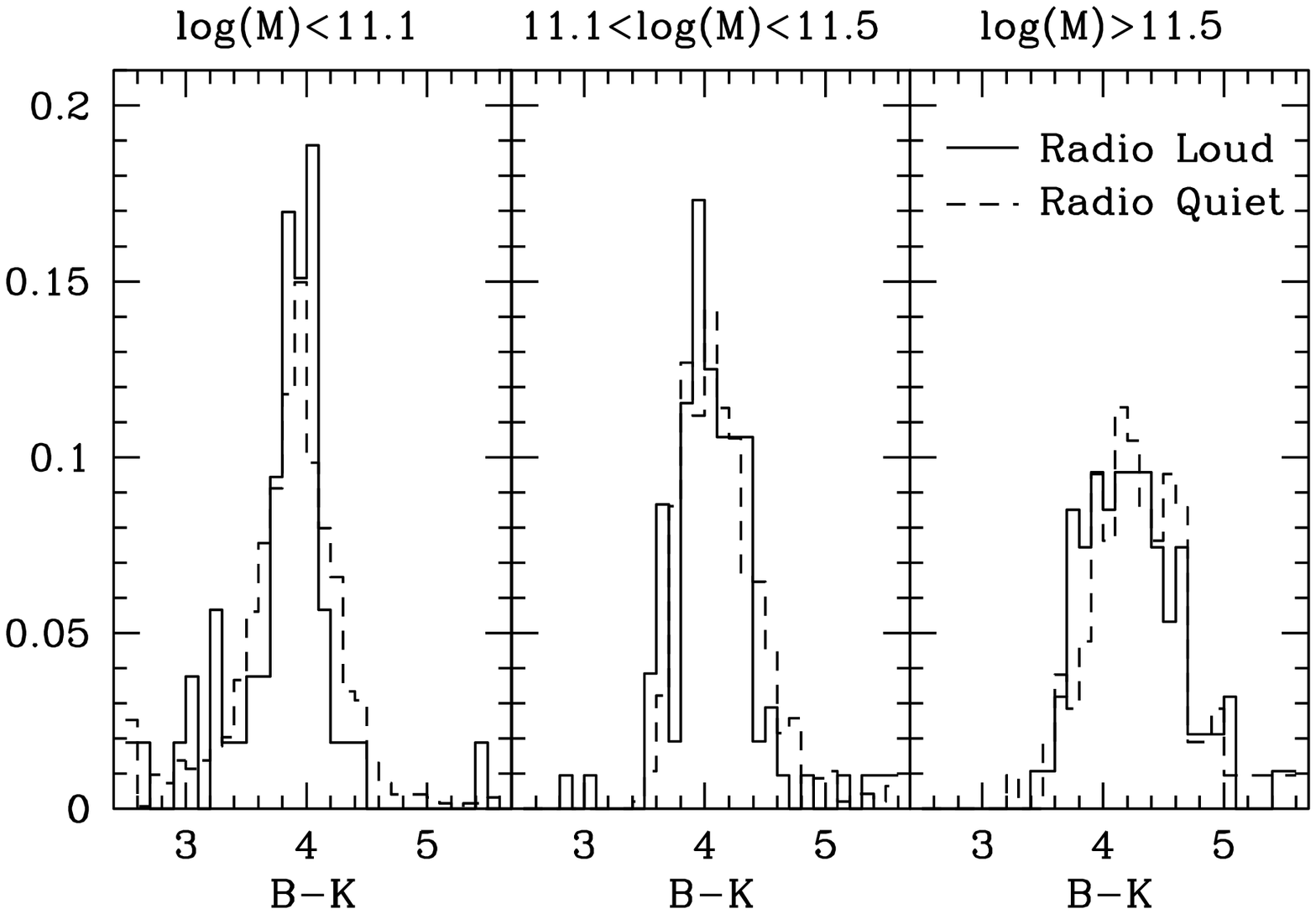}
  \includegraphics[height=6.3cm]{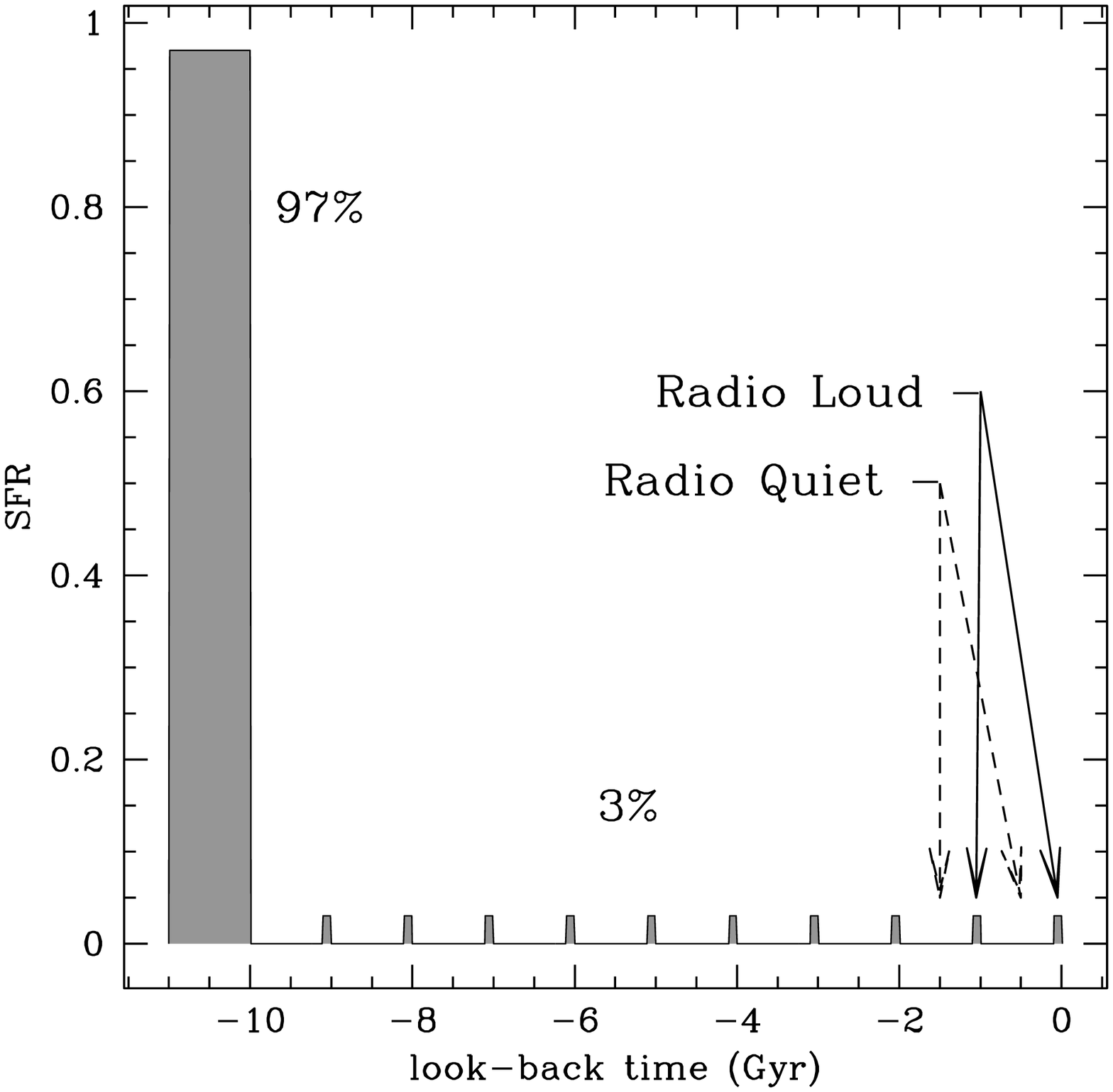}
  \caption{
  {\em Left}: (B--K) color distribution of early-type radio-loud 
  (solid line) and
  radio-quiet galaxies (dashed line) in three stellar mass ranges. The two
  groups of galaxies have practically indistinguishable color distributions, 
  meaning that the stellar populations are similar. 
  {\em Right:} Model of early-type galaxies reproducing both the dichotomy
  radio-loud/radio-faint and the similar (B--K) colors.
  }
\end{figure}

The difference between radio-loud and radio-quiet galaxies can
be reproduced by the model of early-type galaxy shown in the right panel
of figure~2: most of the stars are
formed in a remote past, about $10^{10}$ years ago, while a small minority
of stars are created in a number of subsequent bursts. A galaxy appears
radio-loud when is observed during the burst, radio-faint soon after,
and radio-quiet during the quiescent inter-burst period.
The abundance ratio between radio-quiet and radio-loud galaxies,
about 0.1 in our sample, means that the duty cycle
of the burst events is about 10\%. As the duration of the radio-loud phase is
about 10$^8$ years, in 10$^{10}$ years the early-type galaxies are expected to 
have experienced 10 small bursts, i.e., 
1 every 10$^9$ years and lasting for about $10^8$ years.

This model naturally explains the fact that radio-loud and radio-quiet
early-type galaxies have very similar (B--K) color,
a sensitive indicator of star formation and stellar age.
This is shown in the left panel of Fig.~2, where the two
color distributions are compared.
Only a small difference 
in the median of the two distributions might be present at any mass, i.e.,
the radio-loud galaxies appear to be 0.03-0.06 mag bluer, 
and this could be the effect of last on-going burst of star formation.

The amount of mass in younger stars 
can be estimated from the (B--K) color, that is  consistent
with the value of (B--K)$\sim$4.1 typical of old stellar populations.
By using the \citet{bruzual03} model, we obtain that no more than
3\% of stellar mass can be created in the 10 bursts (0.3\% of mass each)
if we assume negligible extinction, 
otherwise the predicted color would be too blue.
The maximum mass in new stars can reach 5\% 
assuming an average extinction of the new component of $A_V=1$. 
More details will be given in a forthcoming paper.

This model predicts that traces of small amounts of recent star formation 
should be present in most of the local early-type galaxies. This is actually
the case: most of them show very faint emission lines (\citet{sarzi06}),
tidal tails (\citet{vandokkum05}), dust lanes (\citet{colbert01}),
HI gas (\citet{morganti06}), molecular gas (\citet{welch03}), and
very blue UV colors (\citet{schawinski07}).

Using this model with a total fraction of new stars of 3\%, we derive the
results shown in figure~3. 
We see that the theoretical models by \citet{greggio83} and 
\citet{matteucci01}, while giving a
good description of the rates displayed in figure~1, 
predicts too few SNe in the first
$10^8$ years (about 11\%) to accurately fit figure~3.
 The observed rates can be reproduced only by
adding a ``prompt'' component (in this case modeled in terms 
of an exponentially
declining distribution with $\tau=$0.03~Gyr) to a ``tardy'' component (an other
declining exponential with $\tau=$3~Gyr), each one comprising 50\% of the total
number of events.

It should be noted that this {\em strong} bimodality is based on a small number
of SNe (21) in early-type galaxies, and the results of oncoming larger SN
searches are needed to  confirm (or discard) this result.

\begin{figure}   
  \includegraphics[height=5.8cm]{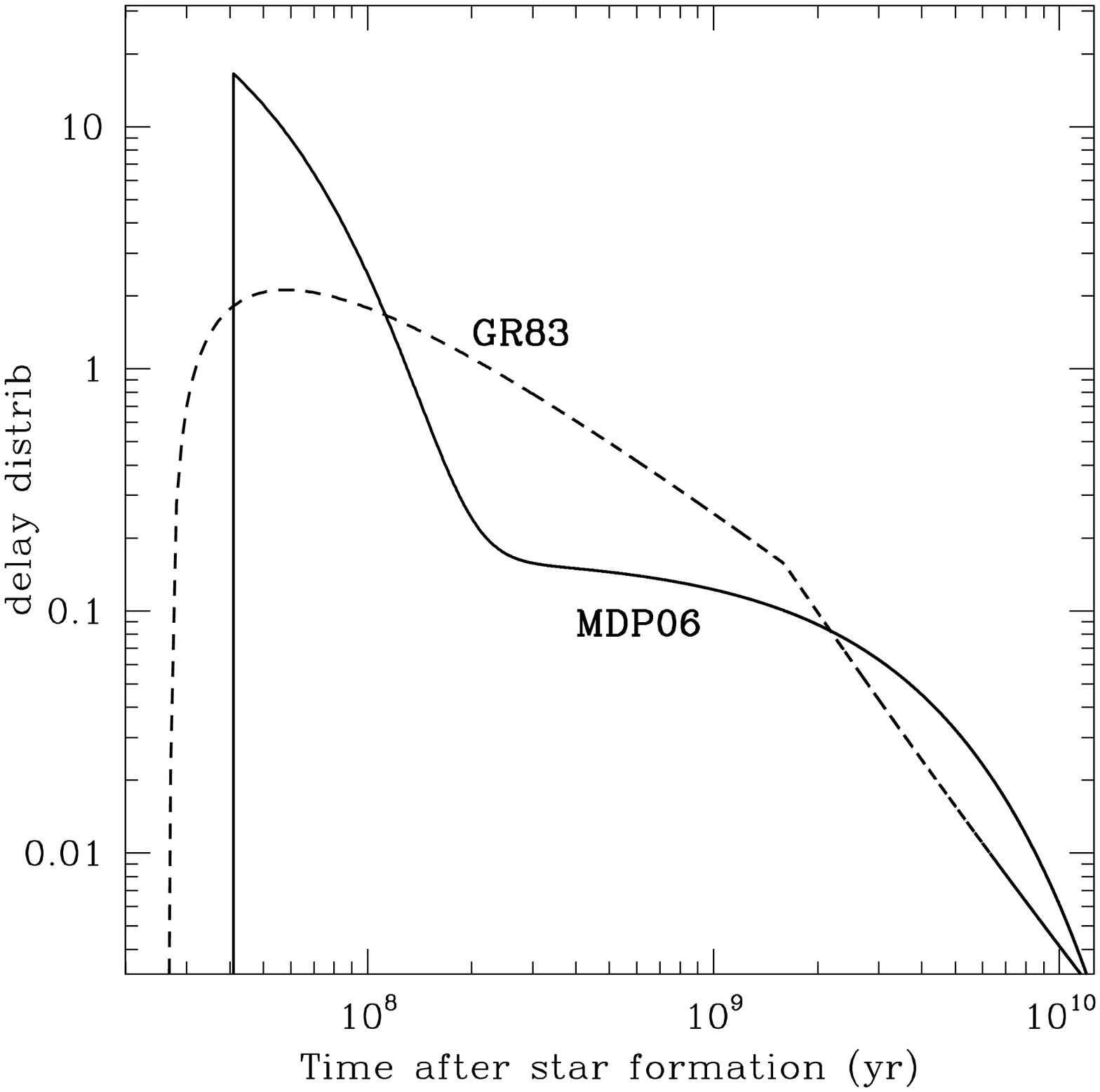}
  \includegraphics[height=5.8cm]{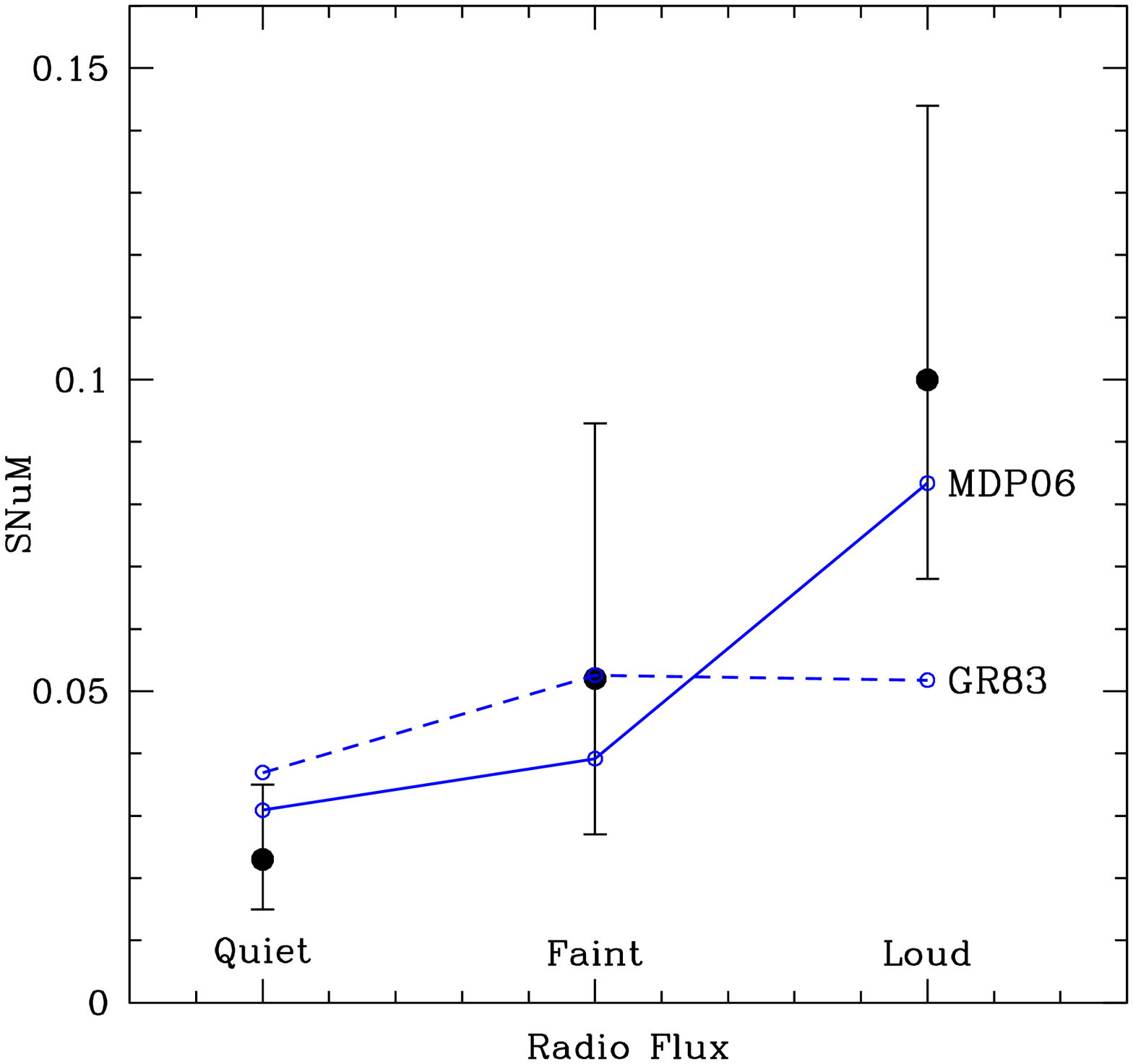}
  \caption{
  {\em Left:} The two DTD studied here, from \citet{greggio83} (GR83)
  and \citet{mannucci06} (MDP06). 
  The latter is the sum of two exponentially declining
  distributions with 3 and 0.03 Gyr of decay time, respectively, each one
  containing 50\% of the events.
  {\em Right:} the solid dots with error bars show the type Ia SN rate as a 
  function of the radio power of the parent galaxy. The dashed line shows
  the results of the GR83 model, the solid one those of MDP06.
  }
\end{figure}

\section{Evolution of the SN rate with redshift}

A related issue is how the rates measured in the local universe
and discussed above are expected to evolve with redshift.
The usual approach is to start from the integrated cosmic star formation 
history and obtain the rates by using some assumptions on progenitors
(for core-collapse SNe) and
on explosion efficiency and DTD 
(for SN Ia, see \citet{mannucci05} for a discussion).
Near-infrared and radio searches for core-collapse supernovae
in the local universe 
(\citet{maiolino02}, \citet{mannucci03}, \citet{lonsdale06})
have shown
that the vast majority of the events occurring in massive starbursts
are missed by current optical searches because they explode in
very dusty environments. 
Recent mid- and far-infrared observations (see \citet{perez05}
and references therein) have shown that the fraction of 
star-formation activity that takes place in very luminous 
dusty starbursts sharply increases with redshift and becomes the dominant 
star formation component at z$\ge$0.5. 
As a consequence, an increasing fraction of SNe are expected to
be missed by high-redshift optical searches. 
By making reasonable assumptions on the number of SNe that can be observed by
optical and near-infrared searches in the different types of galaxies
(see \citet{mannucci07} for details)
we obtain the results shown in figure~4. We estimate that 
5--10\% of the local core-collapse (CC) SNe are out of reach of the
optical searches.
The fraction of missing events rises sharply toward z=1, where about 
30\% of the CC SNe will be undetected. At z=2 the missing 
fraction will be about 60\%. 
Correspondingly, for type Ia SNe, our computations provide 
missing fractions of 15\% at z=1 and 35\% at z=2.
Such large corrections are crucially important to compare
the observed SN rate with the expectations from the
evolution of the cosmic star formation history, 
and to design the future SN searches at high redshifts.

\begin{figure}   
  \includegraphics[height=7.8cm,angle=270]{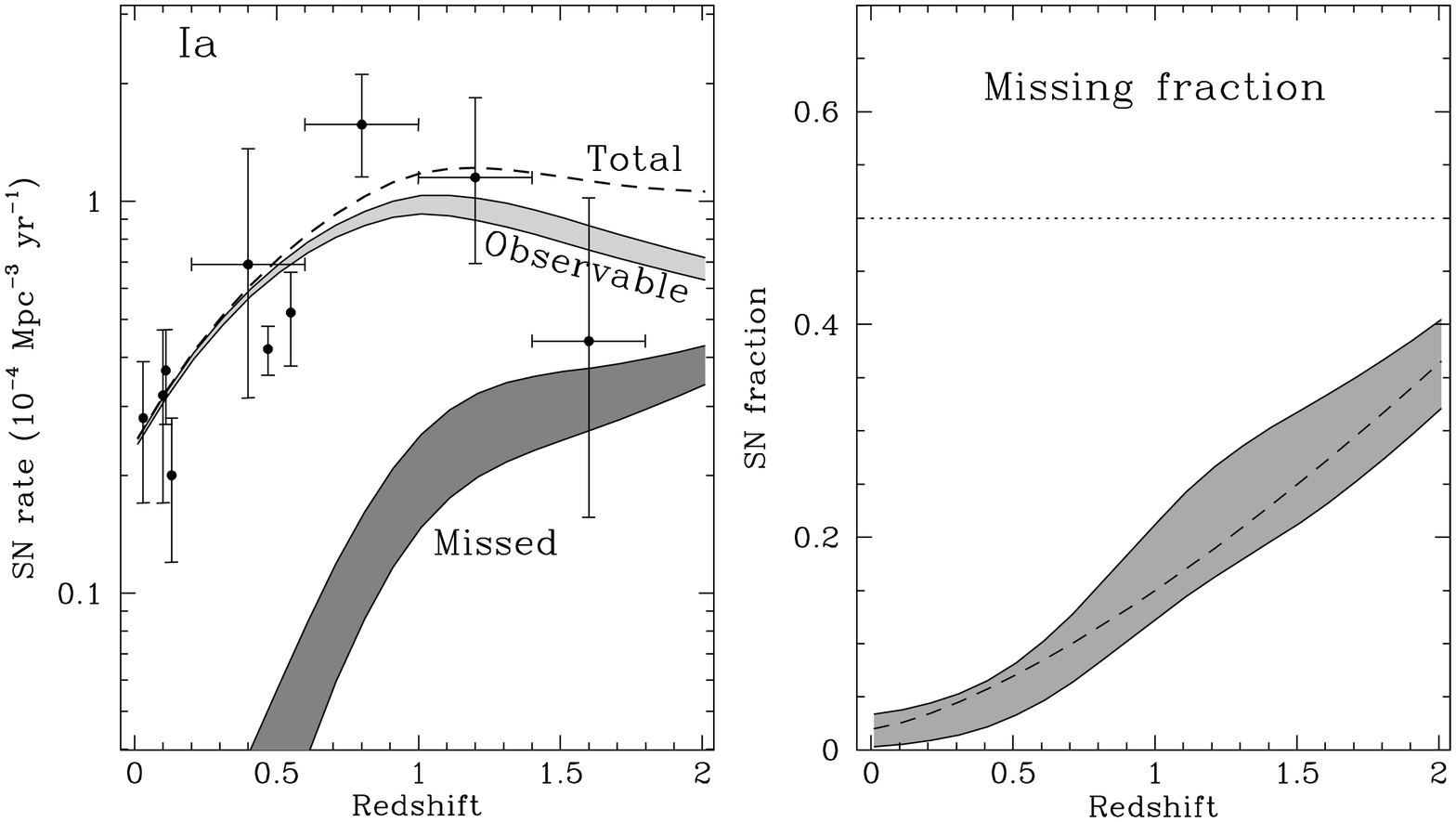}
  \includegraphics[height=7.8cm,angle=270]{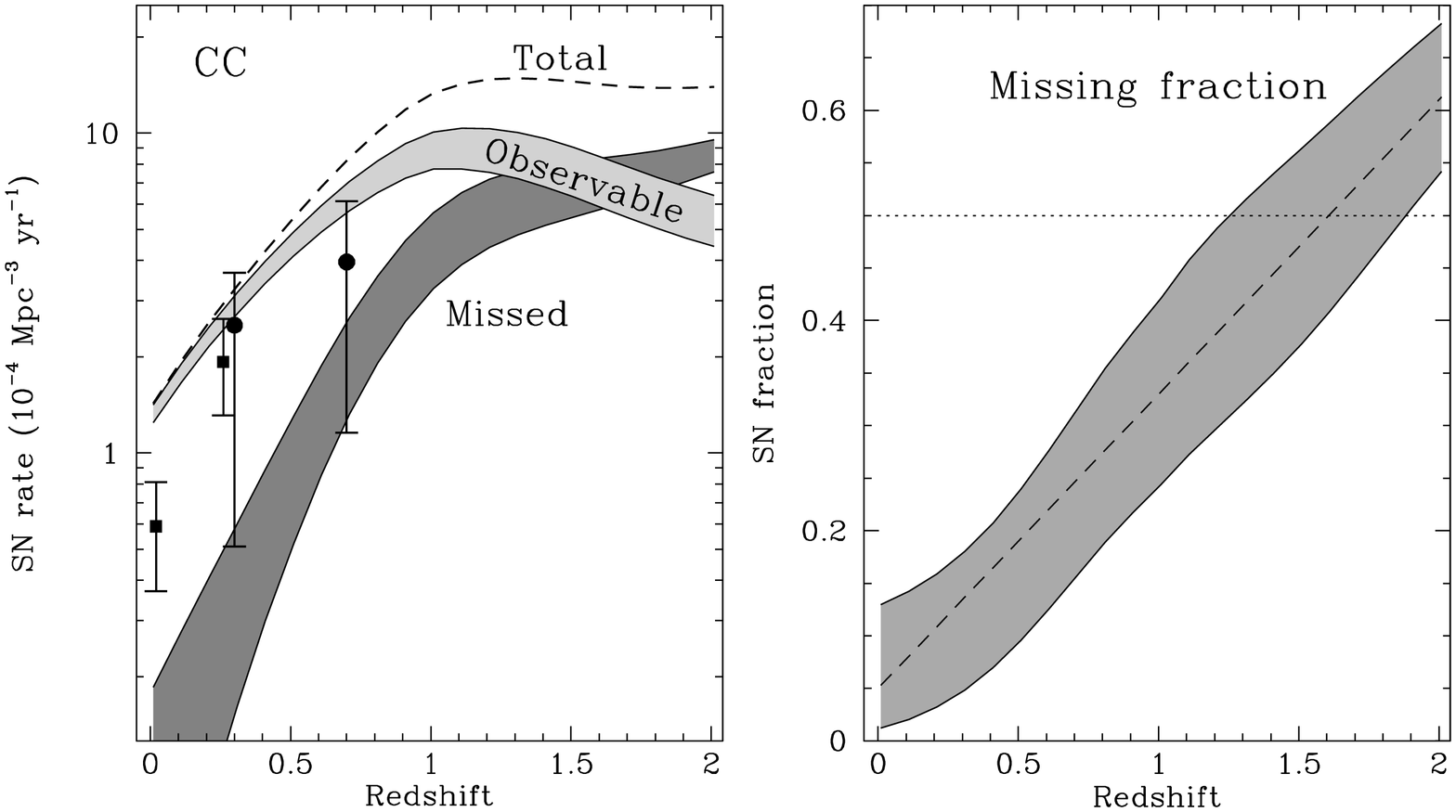}
  \caption{
  Evolution of the rates of type Ia (two left-most panels) 
  and core-collapse SNe (two right-most panels), from
  \citet{mannucci07}. In the first and third panels,
  the dashed line shows the total rate
  expected from the cosmic star formation history, the light grey area
  the rate of SNe that can be recovered by the optical and near-IR searches,
  and the dark grey area the rate of SNe exploding inside dusty
  starbursts and which will be missed by the searches.
  The second and forth panels show the fraction of missed SNe.
  }
\end{figure}





\bibliographystyle{aipproc}   

\bibliography{sample}

\begin{thebibliography}{20}

\bibitem[Minkowski (1941)]{minkowski41}
	R. Minkowski, 1941, \emph{PASP}, 53, 224
\bibitem[van den Bergh (1959)]{vandenbergh59}
    S. van den Bergh, 1959, \emph{AnAp}, 22, 123
\bibitem[Greggio \& Renzini (1983)]{greggio83}
    L. Greggio \& A. Renzini, 1983,  \emph{ApJ}, 118, 217
\bibitem[Mannucci et al. (2005)]{mannucci05}
    F. Mannucci, et al., 2005, \emph{A\&A}, 433, 807
\bibitem[Sullivan et al. (2006)]{sullivan06}
    M. Sullivan et al., 2006, \emph{ApJ}, 648, 868
\bibitem[Scannapieco \& Bildsten (2005)]{scanna05}
    E. Scannapieco \& L. Bildsten, 2005, \emph{ApJ}, 629, L85
\bibitem[Matteucci et al. (2006)]{matteucci06}
    F. Matteucci et al., 2006, \emph{MNRAS}, 372, 265
\bibitem[Calura et al. (2007)]{calura07}
    F. Calura, F. Matteucci, \& P. Tozzi, 2007, MNRAS, in press
	(astro-ph/0702714)
\bibitem[Mannucci et al. (2006)]{mannucci06}
    F. Mannucci, M. Della Valle \& N. Panagia, 2006, \emph{MNRAS}, 370, 773
\bibitem[Greggio (2005)]{greggio05}
    L. Greggio, 2005, \emph{A\&A}, 441, 1055
\bibitem[Della Valle et al. (2005)]{dellavalle05}
    M. Della Valle et al., 2005, \emph{ApJ}, 629, 750
\bibitem[Bruzual \& Charlot (2003)]{bruzual03}
    G. Bruzual \& S. Charlot, 2003, \emph{MNRAS}, 341, 33
\bibitem[Sarzi et al. (2006)]{sarzi06}
    M. Sarzi et al., 2006, \emph{MNRAS}, 366, 1151
\bibitem[van Dokkum (2005)]{vandokkum05}
    P. van Dokkum, 2005, \emph{AJ}, 130, 264
\bibitem[Colbert et al. (2001)]{colbert01}
    J. W. Colbert et al., 2001, \emph{AJ}, 121, 808
\bibitem[Morganti et al. (2006)]{morganti06}
    R. Morganti et al., 2006, \emph{MNRAS}, 371, 157
\bibitem[Welch \& Sage (2003)]{welch03}
    G. A. Welch \& L. J. Sage, 2003, \emph{ApJ}, 584, 260
\bibitem[Schawinski et al. (2007)]{schawinski07}
    K. Schawinski et al., 2007, \emph{ApJ}, in press (astro-ph/0601036)
\bibitem[Matteucci \& Recchi (2001)]{matteucci01}
    F. Matteucci \& S. Recchi, 2001, \emph{ApJ}, 558, 351
\bibitem[Maiolino et al. (2002)]{maiolino02}
    R. Maiolino et al., 2002, \emph{A\&A}, 389, 84
\bibitem[Mannucci et al. (2003)]{mannucci03}
    F. Mannucci et al., 2003, \emph{A\&A}, 401, 519
\bibitem[Lonsdale et al. (2006)]{lonsdale06}
    C. J. Lonsdale et al., 2006, \emph{ApJ}, 647, 185
\bibitem[P\'erez-Gonz\'alez et al. (2005)]{perez05}
    P. G. P\'erez-Gonz\'alez et al., 2005, \emph{ApJ}, 630, 82
\bibitem[Mannucci et al. (2007)]{mannucci07}
    F. Mannucci, M. Della Valle \& N. Panagia, 2007, \emph{MNRAS}, 
	in press (astro-ph/0702355)

\end{thebibliography}

\IfFileExists{\jobname.bbl}{}
 {\typeout{}
  \typeout{******************************************}
  \typeout{** Please run "bibtex \jobname" to optain}
  \typeout{** the bibliography and then re-run LaTeX}
  \typeout{** twice to fix the references!}
  \typeout{******************************************}
  \typeout{}
 }

\end{document}